\documentclass[doublecol]{epl2} 

\title{Two atomic quantum dots interacting via coupling to BECs}
\shorttitle{Two atomic quantum dots interacting via coupling to BECs} 

\author{Anna Posazhennikova\inst{1,2} \and Wolfgang Belzig \inst{1}}

\shortauthor{A. Posazhennikova and W. Belzig}

\institute{                    
 \inst{1}   Fachbereich Physik, Universit\"at Konstanz, D-78457, Konstanz, Germany\\
 \inst{2} Physikalisches Institut, Universit\"at Bonn, Nussallee 12, D-53115, Bonn, Germany
  }


\abstract{ We consider a system of three weakly coupled Bose-Einstein
  condensates and two atomic quantum dots embedded in the barriers
  between the condensates.  Each dot is coupled to two neighboring
  condensates by optical transitions and can be described as a
  two-state system, or a pseudospin 1/2. Although there is no direct
  coupling between the dots, an effective interaction between the
  pseudospins is induced due to their coupling to the condensate
  reservoirs. We investigate this effective interaction, depending on
  the strengths of the dot-condensate coupling $T$ and the direct
  coupling $J$ between the condensates. In particular, we show that an
  initially ferromagnetic arrangement of the two pseudospins stays
  intact even for large $T/J$. However, antiferromagnetically aligned
  spins undergo peculiar ``breathing'' modes for weak coupling
  $T/J<1$, while for strong coupling the behaviour of the spins
  becomes uncorrelated.}

\begin{document}

\maketitle

\section{Introduction}

Atomic quantum dots (AQDs) are a relatively new topic, referred to as
nanobosonics in the context of cold atoms. An AQD constitutes a lowest
atomic level of a tight trapping potential which can be maximally
singly occupied due to a large repulsive interaction energy between the
atoms. In analogy to nanoelectronics the question arises whether one
can couple an atomic dot to large bosonic reservoirs, i.e. ``leads''
and investigate particle transport properties in such a combined
system. Another interesting problem is the control and manipulation of
the single spin behaviour, since such a quantum dot can be described
in terms of a two-state system (the occupied state is equivalent to a
``spin-up'', and the empty state corresponds to a ``spin-down'').

\begin{figure}
\onefigure[width=0.42\textwidth]{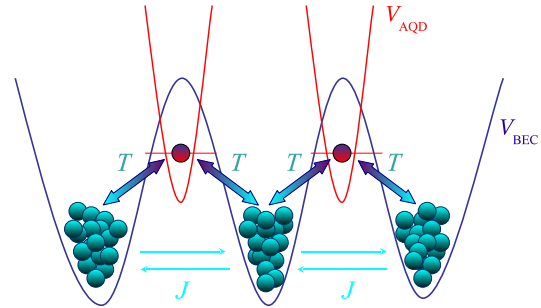}
  \caption{System of two atomic quantum dots coupled to three
    Bose-Einstein condensates in a three-well potential $V_{\rm
      \scriptscriptstyle BEC}$. Inside the barriers between the wells
    atomic quantum dots (single atoms of a different from condensate
    bosons hyperfine species) are embedded. $T$ is the optical
    coupling between the dots and condensate, and $J$ is the Josephson
    tunneling between the wells. }
\label{fig.1}
\end{figure} 

An interesting variant of the optical coupling between an AQD and a
uniform superfluid reservoir was considered by Recati {\it et al}
\cite{Recati}. Provided the dot atoms and the atoms of the superfluid
reservoir are of different hyperfine species, one can (by an external
laser) induce Raman transitions between dot atoms and those of the
condensate. At low energies the spin degrees of freedom couple to the
lowest excitation modes of a superfluid bath - phonons, and therefore
the combined system can be mapped onto a spin-boson model, and the
dissipative behaviour of the spin can be explored \cite{Recati}. From
quantum optical point of view this kind of Raman coupling and its
consequences were numerically investigated by Zippilli and Morigi
\cite{Morigi}.

In previous works by Bausmerth {\it et al.} \cite{Bausmerth} and
Fischer {\it et al.} \cite{Rapid}, a single AQD coherently coupled by
optical transitions to {\it two} BEC-reservoirs with finite number of
particles has been considered. A direct Josephson tunneling between
the two BECs has also been taken into account.  One should note here,
that a Bose Josephson junction exhibits an unusually rich dynamical
behaviour \cite{Smerzi,Milburn,Zapata}, which was also confirmed
experimentally \cite{Albiez}.
  
In previous works \cite{Bausmerth} and \cite{Rapid} it was observed,
that one should distinguish between the two important limiting cases:
(i) the weak coupling regime, in which the coupling between dot and
condensate $T$ is smaller or of the same order as the direct Josephson
coupling $J$ between two condensates, and (ii) the strong coupling
regime, in which the coupling between dot and condensate is
dominating, so that the tunneling between the two condensates occurs
predominantly through the dot channel \cite{Rapid}.

\begin{figure}
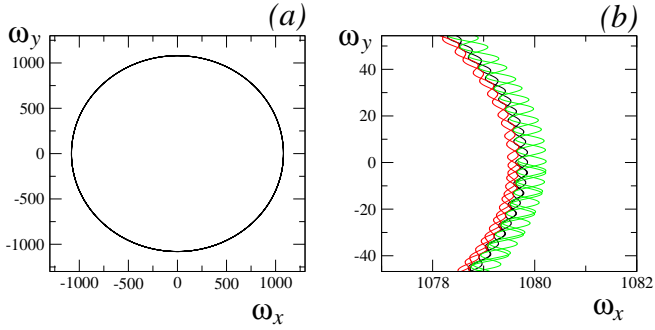

\onefigure[width=0.48\textwidth]{Figure_2.eps}
  \caption{Rotation frequencies (Eqs.~(\ref{freq_left}) and
    (\ref{freq_right})) of the pseudospins for symmetric initial
    conditions in the case of weak coupling. We took
    $N_1(0)=N_3(0)=500$, $N_2(0)=1000$, $U=0$ and the relative
    coupling of the dot to the condensate is $T/J=10$. The
    frequencies $\omega_x^j$ and $\omega_y^j$ are expressed in units
    of $2J$. It is clear that in both FM and AFM cases the spins
    nutate.  Plot (b) shows a zoom in details of plot (a) on a scale,
    on which the differences between the three curves in plot (a) can
    be distinguished. Red and black curves show the frequencies of the
    two initially antiferromagnetically aligned spins, while the green
    line is for initially ferromagnetically aligned spins. The
    detuning energy is fixed to $\hbar \delta/2J=1$.  }
\label{fig2}
\end{figure}

Since in the weak coupling regime the effect of the AQD on the
tunneling between the condensates is very small, it is the favorable
regime in order to investigate the behaviour of the pseudospin
under the influence of an effective time-dependent field, induced by
the condensates dynamics.  It turns out, that the pseudospin behaves in
this case as a quantum top and undergoes multi-frequency rotations
depending on the initial conditions and the mean-field interaction
parameter \cite{Bausmerth}.

In the strong coupling limit ($T\gg J$), it was shown that an atomic
quantum dot, surprisingly, can cause large particle imbalance
oscillations between the wells similar to the Josephson effect, which
are characterized by the two frequencies.  For
non-interacting condensates these frequencies could be derived
analytically \cite{Rapid}.

In this work we discuss the three-well condensate with two AQDs
embedded between the wells (fig.~\ref{fig.1}), coupled to the
condensates by optical Raman transitions. We show that, although
the AQDs do not directly interact with each other, their coupling to
the condensates induces an effective coupling between the two
pseudospins, associated with the two dots, which can be studied
numerically. We show below that symmetric initial conditions favour
the ferromagnetic (FM) coupling between the dots, while
antiferromagnetic (AFM) spin alignment does not survive the increasing
coupling $T/J$.  These features are especially clearly observed in the
strong coupling limit.

\section{Model}

\begin{figure}
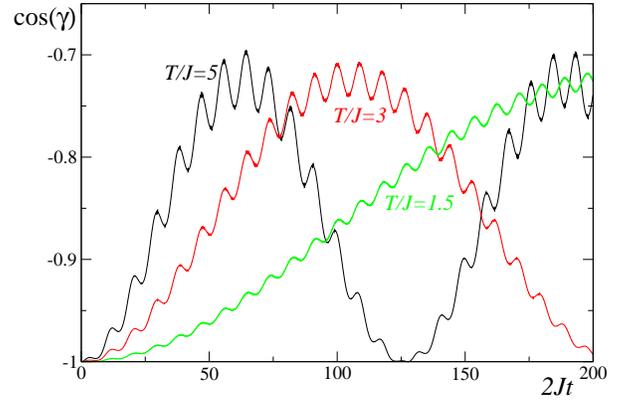

\onefigure[width=0.44\textwidth]{Figure_3.eps}
  \caption{Relative orientiaton $\vec s_1 \vec s_2=\cos(\gamma)$
    versus time for different coupling constant $T/J$. 
    The initial conditions are the same as in
    Fig.~\ref{fig2} and initial alignment of the two pseudospins is
    AFM. One can observe the ``breathing'' modes, whose period
    decreases for growing $T/J$. }
\label{fig3}
\end{figure}

We investigate the setup depicted in Fig.~\ref{fig.1}. The condensate
in a three-well potential $V_{\rm \scriptscriptstyle BEC}$ is
described within a three mode approximation, which is just a
generalization of the two-mode case \cite{Milburn,Smerzi}, so that the
wave-function is $\Psi({\bf r},t)=\sum_{i=1}^3 \phi_i({\bf
  r})\Psi_i(t)$, where the $\phi_i({\bf r})$'s are the normalized local mode
solutions for well $i$, and
\begin{equation}
\Psi_{i}(t)=\sqrt{N_i(t)}e^{i\theta_i(t)}.
\end{equation}
Here $N_i$ is the number of particles in the $i$-th condensate, and
$\theta_i$ is its phase. 

The Hamiltonian of our system reads
\begin{eqnarray}
  H&=&\sum_{i=1}^3E_i
  |\Psi_i|^2+\frac{U}{2}\sum_{i=1}^3|\Psi_i|^4-J(\Psi_1^{\dagger}\Psi_2+\Psi_2^{\dagger}\Psi_1) 
  \nonumber \\ &-&J(\Psi_2^{\dagger}\Psi_3+\Psi_3^{\dagger}\Psi_2)-\hbar
  \delta
  \sum_{i=1}^{2}\frac{1+\sigma_z^{(i)}}{2} \nonumber \\
  &+&T\sum_{i=1,2}\Big[(\Psi_i^{\dagger}+\Psi_{i+1}^{\dagger})\sigma_{-}^{(i)}+h.c.\Big], 
\end{eqnarray}
where $\sigma_{\pm}^{(j)}=(\sigma_x^{(j)}\pm i\sigma_y^{(j)})/2$,
where $\sigma_x^{(j)}$, $\sigma_y^{(j)}$, and $\sigma_z^{(j)}$ are the
Pauli matrices describing the pseudospin degrees of freedom of the
first ($j=1$) or the second ($j=2$) atomic quantum dot. Each quantum
dot is described by a state-vector ${\vec s}^{(j)}=\langle
\Psi_d^j|{\vec \sigma}^{(j)}| \Psi_d^j \rangle $ with $j=1,2$ and
$\Psi_d^j$ being a two-state wave function of the $j$-th dot. This
formalism is applicable in the limit of large interaction on the atomic dots.

For simplicity we consider a symmetric system: the Josephson couplings
between the condensates are the equal: $J_{12}=J_{23}\equiv J$,
defined through
\begin{eqnarray}
  J=-\int d{\bm  r} \Big[\frac{\hbar^2}{2m}(\nabla\phi_i({\bm r})\nabla\phi_{i+1}({\bm r}))
  \nonumber \\+\phi_i({\bm r}) V_{\rm \scriptscriptstyle BEC}({\bm r})
  \phi_{i+1}({\bm r})\Big]\,,\,i=1,2\,.
\end{eqnarray}

\begin{figure}
\onefigure[width=0.44\textwidth]{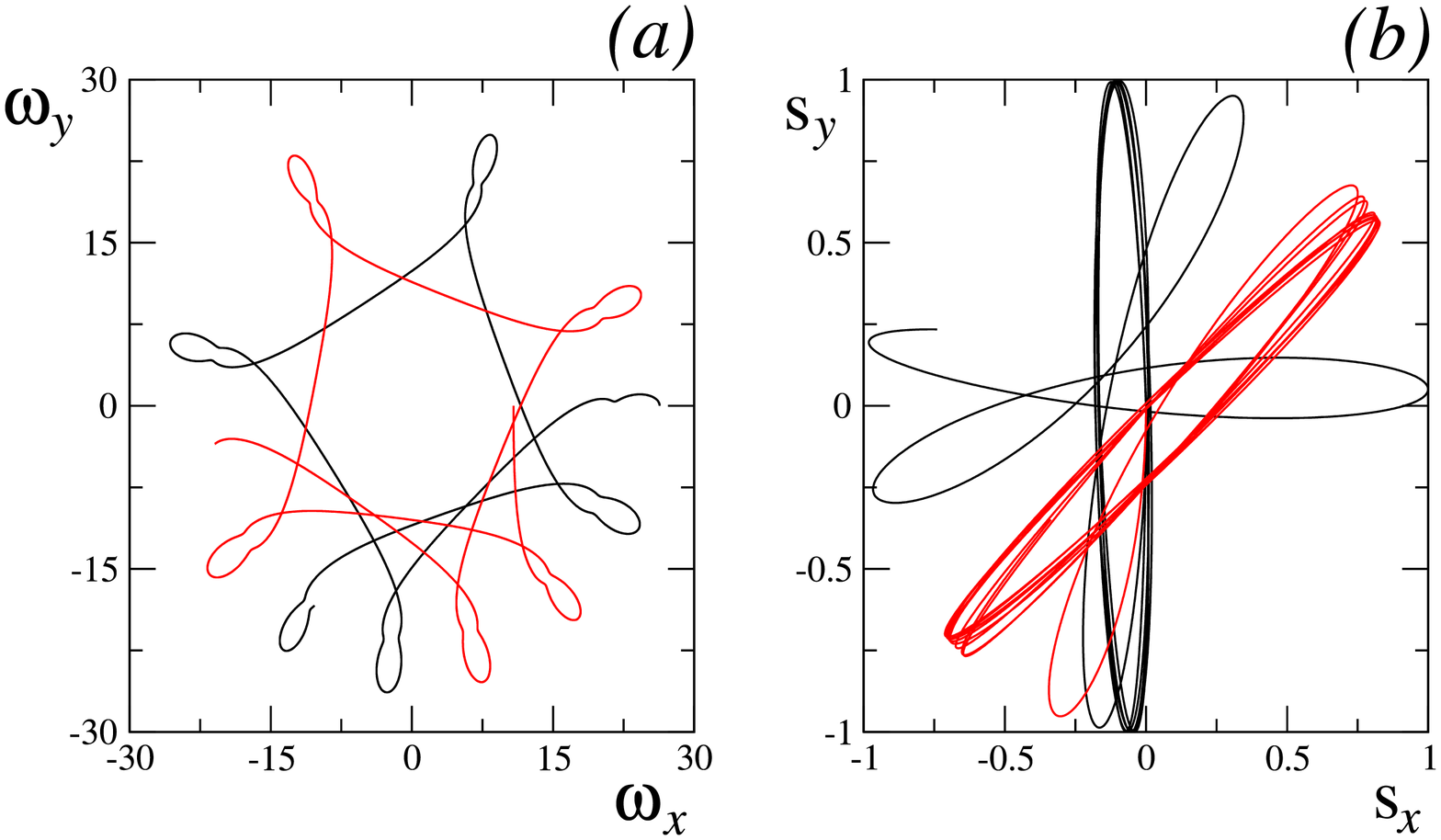}
  \caption{Behaviour of the pseudospins for asymmetric initial
    condensate sizes in the case of weak coupling: $N_1(0)=10000$,
    $N_2(0)=1000$, $N_3(0)=500$, interaction is fixed to $UN_0/2J=1$,
    and the relative coupling of the dot to the condensate is
    $T/J=0.1$. Plot (a) displays the frequency evolution in the
    $\omega_{x}$-$\omega_y$-plane, while in plot (b) the behaviour of
    the pseudospins is shown. The spins were initially FM
    aligned. The black curve denotes the left pseudospin, and red curves
    correspond to the right pseudospin.  }
\label{fig4}
\end{figure}

All four dot-condensate couplings are expressed by a single parameter
$T$
\begin{equation}
T=\hbar \Omega_R\int d{\bm r}\, \phi_j({\bm  r})\phi_d^j({\bm r})
=\hbar \Omega_R\int d{\bm r}\, \phi_{j+1}({\bm r})\phi_d^j({\bm r}),
\end{equation}
where $j=1,2$, and the spatial wave function of each dot
$\phi_d^j({\bm r})$ is normalized to unity.  $\Omega_R$ is the Rabi
frequency of the  Raman transition.
The Rabi frequency is an externally controllable parameter, that can
be tuned to any desirable value.  Spontaneous emission is suppressed
by a large detuning from the excited electronic states, which is
absorbed into the effective dot energy $\hbar\delta$
\cite{Recati}. Another important parameter in the problem is the
standard mean-field two-particle interaction between the condensate
atoms
\begin{equation}
U=g\int d{\bm
  r}\,|\phi_{i}({\bm r})|^4, 
\end{equation}
where $g=4\pi a_s/m$ with $a_s$ being the $s-$wave scattering rate.
The zero-point energies are taken equal to each other $E_i
=\int\left[-\frac{\hbar^2}{2m}\mid\nabla\phi_{i}({\bm
    r})\mid^2+|\phi_{i}({\bm r})|^2V_{\rm \scriptscriptstyle BEC}({\bm
    r})\right]d{\bm r}\equiv E $, and in the actual calculation we
assume $E=0$.

We can now derive equations of motion for the condensates wave
functions, which take the following form
\begin{eqnarray}
\label{eq:schr1}
i\partial_t \Psi_1&=&(E+U|\Psi_1|^2)\Psi_1-J \Psi_2+T s_{-}^{(1)}\,, \\
i\partial_t \Psi_2&=&(E+U|\Psi_2|^2)\Psi_2-J (\Psi_1+ \Psi_3) \nonumber \\
\label{eq:schr2}
&+&T(
s_{-}^{(1)}+
s_{-}^{(2)})\,, \\
\label{eq:schr3}
i\partial_t \Psi_3&=&(E+U|\Psi_3|^2)\Psi_3-J \Psi_2+T
s_{-}^{(2)}\,.
\end{eqnarray}
The equations of motion for the pseudospin components of the two
quantum dots are expressed as two Bloch equations
\begin{equation}
\hbar \partial_t {\vec s}^{(j)}={\vec \omega}^{(j)} \times {\vec s}^{(j)}\,.
\end{equation}
The Bloch equations for the two spins are coupled through the rotation frequencies
\begin{equation}
{\vec \omega}^{(1)}=\left( \begin{array}{c}
2T (\Re \Psi_1 +\Re \Psi_2) \\
-2T (\Im \Psi_1 +\Im \Psi_2) \\
-\hbar \delta \end{array}
\right),
\label{freq_left}
\end{equation}
and
\begin{equation}
{\vec \omega}^{(2)}=\left( \begin{array}{c}
2T (\Re \Psi_2 +\Re \Psi_3) \\
-2T (\Im \Psi_2 +\Im \Psi_3) \\
-\hbar \delta \end{array}
\right),
\label{freq_right}
\end{equation}
which are time-dependent due to a the coupling to the condensate
dynamics Eqs.~(\ref{eq:schr1})-(\ref{eq:schr3}).

We solve this set of equations numerically for different initial conditions,
and analyze the behaviour of the condensates and the
dots in both regimes of weak and strong coupling. In case of the condensate we
are interested in the behaviour of normalized particle imbalances
\begin{equation}
n_{12}(t)=\frac{N_1(t)-N_2(t)}{N_0}\,, \quad n_{23}=\frac{N_2(t)-N_3(t)}{N_0},
\label{imb}
\end{equation} 
where $N_0=N_1(0)+N_2(0)+N_3(0)$. Note, that conserving quantities are
$N_{tot}=N_0+(s_z^{(1)}+1)/2+(s_z^{(2)}+1)/2$ and
$(\vec s^{(j)})^2=(s_x^{(j)})^2+(s_y^{(j)})^2+(s_z^{(j)})^2$ for $j=1,2$.

\section{Weak coupling limit}

\begin{figure}
\onefigure[width=0.48\textwidth]{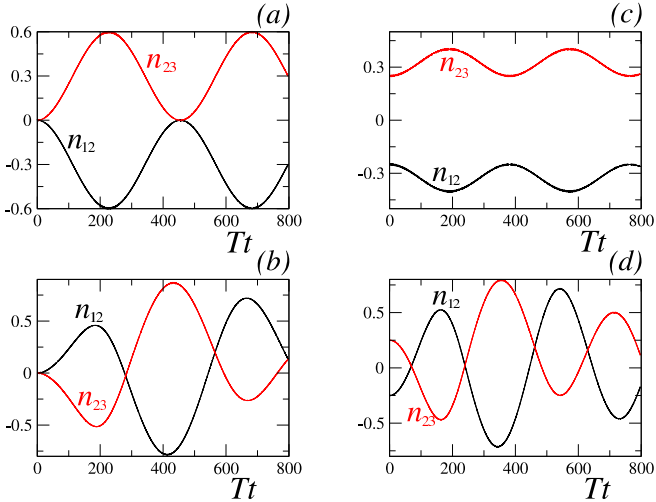}
\caption{Example of large amplitude particle oscillations between the wells
  $n_{12}$ and $n_{23}$ in case of strong coupling
  limit. Left column (a) is for initial conditions
  $N_1(0)=N_2(0)=N_3(0)=1000$, right column (b) is for initial conditions
  $N_1(0)=N_3(0)=500$, $N_2(0)=1000$. For all plots the effective interaction
  $UN_0/T=0.01$ and the detuning $\delta/T=100$. Upper plots are for initial FM
  arrangement of the two AQDs, and lower plots are for initially AFM dots.}
\label{fig5}
\end{figure}

The weak coupling limit implies that $T\le J$ (we note that, since a
quantum dot can be only singly occupied, $T$ can be also order of
magnitude larger than $J$, but we still refer to this realization as
the weak coupling limit).  In this limit the effect of the AQDs on the
dynamics of the Josephson tunneling between the condensates is
expected to be insignificant \cite{Bausmerth}. We concentrate
therefore on the pseudospin behaviour.

It turns out, that symmetric initial conditions, which means that the
 two outer wells are equally populated
$N_1(0)=N_3(0)$, favour nutations of the two pseudospins
(Fig.~\ref{fig2}(a)), which remain parallel if they were initially
FM aligned, and antiparallel if they were AFM
aligned at $t=0$. This is true for a very weak coupling $T\ll J$, 
when the effective interaction between the spins is
negligibly small.

However, with increasing coupling, a certain correlation between the
two pseudospins develops, and their dynamical behaviour starts being
dependent on their initial alignment. These features are to be
observed in Fig.~\ref{fig2}(b), in which part of
Fig.~\ref{fig2}(a) is enlarged. The two FM spins remain parallel and
their frequencies follow the same (green) curve. However, initially
antiparallel spins do not remain antiparallel with time. The red curve
in fig. \ref{fig2}(b) corresponds to the left dot, while the black
curve corresponds to the right dot.

In order to understand better this behaviour we analyze the time evolution of
the relative angle $\gamma$ between the spins, defined through
\begin{equation}
\vec s_1 \vec s_2=\cos\gamma.
\end{equation}
The time-dependence of $\cos\gamma$ for different coupling constants
$T/J$ is shown in Fig.~\ref{fig3}. The angle $\gamma$ oscillates
between $\pi$ and $\sim 0.75 \pi$. In the following we refer to these
oscillations as ``breathing'' modes. The breathing modes become slower
for smaller coupling $T/J$ and eventually disappear for $T\ll
J$. We conclude that the breathing modes are indications of the
induced correlation between the spins.

One can also consider asymmetric initial conditions: $N_1(0)\neq
N_2(0)\neq N_3(0)$ as, for instance, in Fig.~\ref{fig4}. In this case
the dynamics of the two AQDs becomes very complicated, although not
chaotic, as the Fourier transform of $(\cos\gamma)$ does not produce a
typical white noise signal.

\section{Strong coupling limit}

\begin{figure}
\onefigure[width=0.44\textwidth]{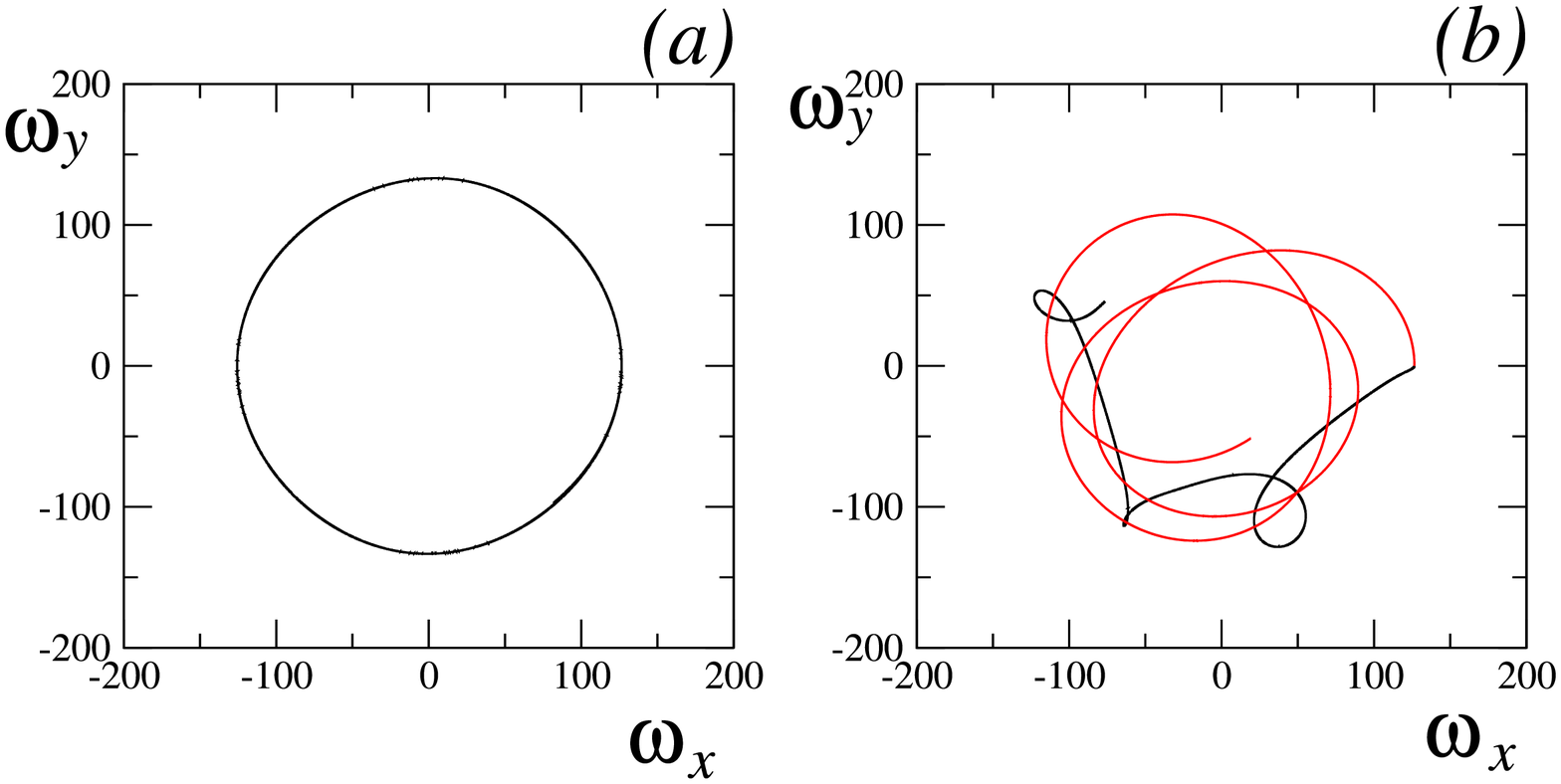}
  \caption{Frequencies $\omega$ for the case of initial conditions of
    Fig.~\ref{fig5}. Initially FM aligned spins remain FM and nutate
    as becomes clear from (a), while AFM aligned spins decohere. The
    black curve is the for left spin and the red curve is for the
    right spin of two initially AFM aligned quantum dots.}
\label{fig6}
\end{figure}

\begin{figure}
\onefigure[width=0.48\textwidth]{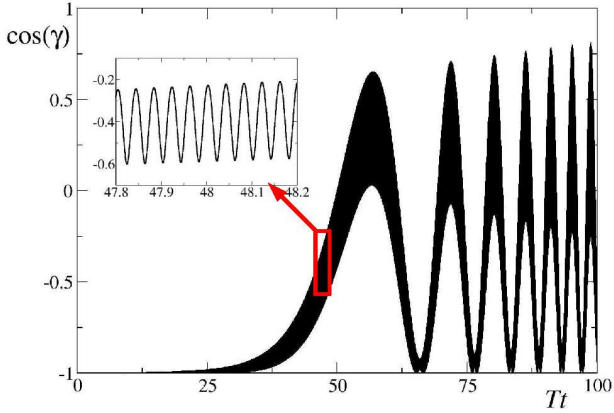}
  \caption{Time-dependence of $\vec s_1 \vec s_2=\cos(\gamma)$ for the
    strong coupling case. Initial conditions are as in
    Fig.~\ref{fig5}(b). }
\label{fig7}
\end{figure}

The peculiarities, which we found out in the dynamics of the initially
AFM aligned spins in the weak coupling limit, are expected to become
even more pronounced in the strong coupling regime, corresponding to
$T \gg J$. In this case, the relevant time scale is set by $T$. It was
previously shown that the ratio $UN_0/\delta$ plays an important role
in the physical behaviour of the system \cite{Rapid}. Thus, the most
interesting regime, when the AQD serves as a quantum ``shuttle'' of
the atoms between the condensates, is achieved for $\delta\gg UN_0$.
Naively one expects, that the oscillations induced solely due to the
coupling of the condensates to the dot, would be always very small,
however, surprisingly large amplitude Josephson-type oscillations
between the wells were obtained \cite{Rapid}.

In our case we also observe this property of AQDs
(Fig.~\ref{fig5}). For initially parallel spins we observe symmetric
oscillations of the particle imbalances $n_{12}$ and $n_{23}$ (see 
Figs.~(\ref{fig5}a) and (\ref{fig5}c)). These are 
 similar to the oscillations in a triple well potential which are always
 symmetric for symmetric initial conditions.
Note, that these
oscillations are induced purely by the dots.

The pseudospins remain parallel, independent of the coupling strength
between them, so that $\cos\gamma(t)=\cos\gamma(0)=1$. The frequency
$\vec \omega$ (which is the same for left and right spin) exhibits
precession (Fig.~\ref{fig6}(a)), which means that each of the
pseudospins nutates.

The situation is very different for initially antiparallel
pseudospins. First of all, the dynamics of interwell tunneling is of
different character and seems to be not very sensitive to initial
conditions for $N_i(0)$ (Figs.~(\ref{fig5}b) and
(\ref{fig5}d)). Second of all, one clearly observes asymmetric
features, which we associate with the uncorrelated spin
behaviour. This can be also seen from the behaviour of $\vec
\omega^{(1)}$ and $\vec \omega^{(2)}$, which is shown in
Fig.~(\ref{fig6}b). For instance, one can notice, that the right (red)
pseudospins oscillates faster. Finally, the time-dependence of
$\cos(\gamma)$ is depicted in Fig.~(\ref{fig7}), where the uncorrelated
dynamics of the two initially antiparallel ($\cos(\gamma)(t=0)=-1$)
pseudospins is presented. For small-amplitude oscillations the
qualitative behaviour of the two AQDs is similar.

\section{Conclusions and Discussion}

We have discussed the problem of an induced interaction between two
atomic quantum dots which are embedded between three trapped
Bose-Einstein condensates. As it is a multi-parameter problem, it
helps to distinguish two different regimes: a weak coupling limit
(when the direct Josephson coupling between the condensates is dominating)
and a strong coupling case, in which the tunneling between the
condensates occurs predominantly through the additional channels
mediated through the dots.

Our main conclusion is that indeed we clearly observe the signs of an
effective interaction, induced between the two dots due to their
coupling to the same condensate system. For this interaction to
appear, $T$ should be at least of the order of $J$. Interestingly,
this induced interaction produces very different effects on the
pseudospins, depending on whether they were initially FM or AFM
aligned.

It turns out, that initially parallel spins are stable with respect to
coupling between the dots and the condensates. They remain parallel even in
the strong coupling limit, while undergoing two-frequency behaviour, as we can
judge from the precessions of $\vec \omega$.

Initially antiparallel spins perform what we call ``breathing'' motion, which
is characterized by small periodic deviations of the relative spin angle $\gamma$
from $\pi$. The breathing modes, however, do not survive the strong coupling
limit, when the effective interaction between the two spins becomes
large. Instead, spins become rather uncorrelated. 

Our analysis has important consequences for the periodic model of the
system under consideration, which is the subject of future
research. First of all, in a periodic system one shall be able to
locally address and modify the interaction between the neighboring
spins just by changing the occupation of each well. Second of all, by
changing the coupling between AQDs and condensates (or optical lattice
sites), one can control the different phase transitions in such a
coupled system. For example, if an optical lattice was initially in
the Mott state, a finite $T$ should drive the system to a superfluid
state once some critical value is exceeded. At the same time spins of
the AQDs will become correlated, and the question will be if this
correlaion is associated with a phase transition.

In the future it will be definitely worth studying quantum effects in
such a model, as well as temperature effects, which would lead to
decoherence phenomena and temperature-dependent phase transitions.

\acknowledgments A.P. acknowledges U. R. Fischer for discussions
related to this work.  Financial support from DFG via SFB 767 is
acknowledged.

\end{document}